\documentclass[aps,pre,twocolumn,superscriptaddress]{revtex4-1}
\usepackage{epsfig,amsmath,amssymb,amsfonts,graphicx,color,calc,epstopdf}

\usepackage[T1]{fontenc}

\usepackage{xcolor,hyperref}
\hypersetup{
   colorlinks,
   linkcolor={blue!50!black},%{red!80!black},
   citecolor={blue!50!black},
   urlcolor={blue!80!black}
}
\let\a=\alpha \let\b=\beta  
   
\let\l=\lambda

 \def\QQ{{\cal Q}}

   \def\qq{{\bf q}}

\def\de{\mathrm{d}}

\newcommand{\beq}{\begin{equation}} 
\newcommand{\eeq}{\end{equation}}
\newcommand{\ba}{\begin{eqnarray}}
\newcommand{\ea}{\end{eqnarray}}

\begin{document}

\title{Quantum exploration of high-dimensional canyon landscapes}
%\title{Fractal free energy landscape in high-dimensional models of confluent tissues}

\author{Pierfrancesco Urbani}
\affiliation{Universit\'e Paris-Saclay, CNRS, CEA, Institut de physique th\'eorique, 91191, Gif-sur-Yvette, France}

 \begin{abstract}
Canyon landscapes in high dimension can be described as manifolds of small, but extensive dimension, immersed in a higher dimensional ambient space and characterized by a zero potential energy on the manifold. Here we consider the problem of a quantum particle
exploring a prototype of a high-dimensional random canyon landscape. We characterize the thermal partition function and show that around the point where the classical phase space has a satisfiability transition so that zero potential energy canyons disappear, moderate quantum fluctuations have a deleterious effect and induce glassy phases at temperature where classical thermal fluctuations alone would thermalize the system.
Surprisingly we show that even when, classically, diffusion is expected to be unbounded in space, the interplay between quantum fluctuations and the randomness of the canyon landscape conspire to have a confining effect.
\end{abstract}

\maketitle

\paragraph*{Introduction --}
The interplay between quantum fluctuations and glassiness has been extensively investigated in the past, see \cite{cugliandolo2022quantum} for a recent review.
In simple landscapes, one expects that quantum fluctuations reduce glassiness since they open  relaxation channels (quantum tunneling) to overcome energy barriers.
Such mechanism is at the basis of quantum annealing algorithms to solve optimization problems.

However, in \cite{markland2011quantum} it was suggested, based on a numerical approximation scheme for quantum dynamics (Mode Coupling Theory (MCT)), that some supercooled liquids can become glassy due to quantum effects. 
The example considered in \cite{markland2011quantum} was quantum hard spheres. It was suggested that the packing fraction at which hard spheres become glassy is lowered when the de Broglie length is finite but not too large. An intuitive argument for this phenomenon is that when $\hbar$ is switched on, spheres acquire a skin of the order of the de Broglie length due to the Heisenberg principle. Therefore the effective density is higher than the nominal one (fixed by the classical diameter). This induces glassy phases where classical diffusion would thermalize the system.

The same phenomenon was observed in \cite{thomson2020quantum} for a simple thermally isolated glass former in high dimension.
The MCT equations considered in \cite{markland2011quantum} are approximate and uncontrolled for hard spheres in three dimensions but become exact in the case of \cite{thomson2020quantum} and can be integrated numerically. However, to claim for absence of thermalization, one needs to perform a numerical extrapolation of correlation functions at long times.

The analysis of quantum models of hard spheres was started in \cite{franz2019impact} where a high-dimensional prototype model of dense hard spheres, the quantum perceptron, was studied. The model describes a high dimensional quantum billiard: a free particle in a high-dimensional non-convex phase space. The quantum partition function can be analyzed by means of the replica method and the saddle point equations can be derived. A tentative numerical solution of the corresponding equations was presented in \cite{artiaco2021quantum}. However this solution is limited by the fact that the saddle point equations  involve the evaluation of the path integral of an impurity problem which must be obtained through quantum Monte Carlo. Therefore, going to low temperature is very demanding.

In this work we consider a different set of quantum models whose classical version was recently motivated by the study of the statistical mechanics of confluent tissues.
Consider a three dimensional confluent tissue \footnote{Note that everything can be generalize to different spatial dimensions} where cells are squeezed and tassellate the space.
A simple model to describe cell-cell excluded volume and elastic interactions is built by considering
a set of points in space, the centers or nuclei of the cells, and constructing the corresponding Voronoi tessellation which defines the boundaries of the cells.

Then the simplest classical Hamiltonian for such system can be written as
\beq
H= \frac 12 \sum_{i=1}^{N} \left[ \left(A_i-A_0\right)^2+ \left(V_i-V_0\right)^2\right]
\label{tissue}
\eeq
being $N$ the total number of cells and $A_i$ and $V_i$ their corresponding surface area and volume.
Eq.~\eqref{tissue} describes a system of cells competing to reach a target area and volume identified by the constants $A_0$ and $V_0$.
The model in Eq.~\eqref{tissue} was simulated in \cite{merkel2018geometrically} where it was observed that tuning $A_0$ \footnote{Note that since the Voronoi tessellation covers the space, $V_0$ can be set to be the average volume per cell.} one goes from a zero temperature liquid phase where $H$ is zero at large $A_0$, to a glassy phase for small $A_0$
where local minima of $H$ have positive potential energy. 
The most interesting phase for the sake of this work, is the liquid one: there, cells can diffuse without increasing the energy.
This means that the motion in phase space happens on a manifold that is fixed by the equations $A_i=A_0$ and $V_i=V_0$ for each cell in the tissue.
We can interpret this portion of phase space as a \emph{canyon} in the potential energy landscape and therefore we refer to this landscape as a canyon landscape.

The model in Eq.~\eqref{tissue} can be canonically quantized by adding a kinetic term:
\beq
H = \frac 12 \sum_{i=1}^{N} \left[ p_i^2 + \left(A_i-A_0\right)^2+ \left(V_i-V_0\right)^2\right] 
\label{Quantum_tissue}
\eeq
where $p_i$ is the momentum operator which satisfies canonical commutation relations with the operators $x_i$ that denote the positions of cells.
The analytical study of the model described in Eq.~\eqref{Quantum_tissue} is essentially prohibitive and one needs to resort to simulations.
In order to make progress we consider a high-dimensional mean field model that was recently studied in detail \cite{urbani2022continuous} and was shown to have the same features
of the model in Eq.\eqref{tissue}.

\paragraph*{The model --}
One way to think about the Hamiltonian in Eq.\eqref{tissue} at zero temperature, is to regard the area and volume terms as non-linear equality constraints for the positions of the cells.
Furthermore, the total number of such constraints is smaller than the total number of degrees of freedom and this allows to have a phase transition from a liquid to a glass phase at $T=0$.
Therefore one way to make a mean field model of such problem is to consider a set of random non-linear equality constraints in high-dimension.

Therefore, we consider an $N$-dimensional degree of freedom $\underline x = \{x_1,\ldots, x_N\}$. Its quantum dynamics is described by the following Hamiltonian
\beq
H = \frac 12 \sum_{i=1}^N p_i^2 + \sum_{\mu=1}^M v(h_\mu(\underline x)-p_0)
\label{quantum_tis_mean}
\eeq
where the momenta $p_i$ satisfy canonical commutation relations with the positional degrees of freedom $x_i$, $[x_i,p_j]=i\hbar \delta_{ij}$.
The potential energy term is given by $v(h)={z^2}/{2\epsilon}$, being $\epsilon$ an energy scale, 
and the gap functions $h_\mu(\underline x)$ are defined by
\beq
h_\mu(x) = \frac{1}{N} \sum_{i<j} J_{ij}^\mu x_ix_j
\eeq
where the matrices $J^\mu$ are symmetric with random normally distributed entries.
The potential term in Eq.~\eqref{quantum_tis_mean} is given as a sum over $M$ terms and we assume that $M$ scales with $N$ as $M=\alpha N$ with $\alpha$ a control parameter of the problem.
The other control parameter is $p_0$.
We consider two variants of the model: in the first one the degrees of freedom are constrained on a compact phase space defined by a spherical constraint
$|\underline x|^2=N$. In the second variant instead we assume that the coordinates $x_i$ are in ${\mathbb R}^N$.

\paragraph*{The classical phase space --}
The analysis of the classical Gibbs measure can be performed using the replica method \cite{MPV87}.
The spherical model was solved in \cite{urbani2022continuous}, see also \cite{fyodorov2022optimization, tublin2022few} for related work on a similar model. 
At fixed $\alpha<1$ and increasing $p_0$ one goes from a satisfiable (SAT) phase
where at zero temperature it exists a canyon manifold so that $h_\mu=p_0$ for all $\mu=1,\ldots, M$, to an unsatisfiable (UNSAT) phase where there are no local minima of zero potential energy.

Interestingly, since the model is non-convex, the SAT phase contains two regions.
The small $p_0$ region is in a replica symmetric (RS) phase where the canyon can be explored ergodically.
Increasing $p_0$, before reaching the SAT/UNSAT transition, one can find a replica symmetry breaking (RSB) point where the canyon undergoes a RSB instability. 
RSB corresponds to a slowdown of simulated annealing to equilibrate the low energy canyon manifold at zero temperature \cite{MPV87}.
 
The solution of the model without spherical constraint can be easily obtained from \cite{urbani2022continuous}.
We first note that the potential energy term in Eq.~\eqref{quantum_tis_mean} is not invariant under rescaling $\underline x\to\lambda \underline x$. The overall scale of $\underline x$ is expected to be fixed by $p_0$.
However an inspection of the solution of the classical model, shows that for $\alpha>1/2$, the zero temperature classical Gibbs measure is dominated by configurations that have a typical length in phase space given by:
\beq
\frac 1N |\underline x|^2 = p_0\sqrt{\frac{4\alpha}{2\alpha -1}} \ \ \ \alpha>1/2
\eeq
while for $\alpha\to1/2^+$, $|\underline x|^2/N\to \infty$ and the classical potential is not confining.
This can be understood in the following way. If $M=1$, one has that the constraint $h_1=p_0$ is an hyperbolic surface.
Indeed diagonalizing the matrix $J$ and rotating the reference frame along ita eigenvectors one can show that $|\underline x|^2$ can have arbitrarily large values because the matrix $J$ has both positive and negative eigenvalues.
The replica computation shows that this mechanism breaks down at $\a=1/2$.
Furthermore, there is a RSB transition at $\alpha=1$.

\paragraph*{Solution of the quantum models --}
We now consider the full Hamiltonian in Eq.~\eqref{quantum_tis_mean}.
We want to study the partition function $Z$ at inverse temperature $\beta=1/T$ and the corresponding free energy $F_J$ as given by
\beq
Z=\mathrm{Tr}\,  e^{-\beta H} \ \ \ \ F_J= -\frac 1{\b N} \ln Z\:.
\eeq
We are interested in the average behavior of the free energy
\beq
F = \overline{F_J}= -\frac 1{\b N} \overline{\ln Z}
\eeq
and the overline denotes the average over the couplings $J^\mu_{ij}$.
Using the replica method, see \cite{FPSUZ17, franz2019impact},
we can obtain the average free energy through a saddle point computation.
This is given in terms of a set of order parameters that are fixed as the saddle points of the large deviation principle
which naturally leads to a set of \emph{overlap} order parameters defined as
\beq
Q_{ab}(t,t') = \frac{1}{N}\overline{\langle \underline x_a(t)\cdot \underline x_b(t')\rangle }
\eeq
where the time index $t\in [0,\b \hbar]$ is a Matsubara time on the thermal circle.
Following \cite{bray1980replica} one can show that 
\beq
\begin{split}
Q_{aa}(t,t')&= q_d(t-t')\\
Q_{a\neq b}(t,t') &= q_{ab}\:.
\end{split}
\eeq
In other words, $q_d(t)$ is a periodic function of $t$. In the spherical model one has the additional constraint that $q_d(0)=1$.
Instead, the off-diagonal matrix elements of $Q_{ab}$ are independent on the Matsubara time.
It is convenient to define the Fourier representation for $q_d(t)$. 
Denoting by
\beq
\begin{split}
\nu_n&=\frac{\omega_n}{\b \hbar}\ \ \ \ \omega_n=2\pi n\ \  \ \ \ \ n\in {\mathbb Z}
\end{split}
\eeq
the Matsubara frequencies of the system, we have
\beq
\begin{split}
q_d(t)&=\sum_{n\in {\mathbb Z}} \tilde q_d(\nu_n)e^{i\nu_n t} \ \ \ \tilde q_d(\nu_n)=\int_{0}^{\beta \hbar} \frac{\de t}{\b \hbar} q_d(t)e^{-i\nu_n t}
\end{split}
\eeq
We also define the rescaled Matsubara time $\hat t=t/\b\hbar$ and $\tilde \qq_d(\omega_n)\equiv \tilde q_d(\nu_n)$
so that
\beq
\begin{split}
\qq_d(\hat t) = \sum_{n\in {\mathbb Z}} \tilde \qq_d(\omega_n)e^{i\omega_n \hat t}\ \ \ \ \ \tilde \qq_d(\omega_n)&=\int_{0}^{1} \de \hat t \qq_d(\hat t )e^{-i\omega_n \hat t}
\end{split}
\eeq
and $q_d(t) = \qq_d(\beta\hbar \hat t)$.
Finally we assume that the off-diagonal part of the matrix $Q_{ab}$ is parametrized by a Parisi function $\QQ(x)$ or by its inverse $x(\QQ)$,
with $x\in [0,1]$ and $\QQ\in [\QQ_m,\QQ_M]$ \cite{MPV87}.

Both order parameters $\qq_d(\hat t)$ and $\QQ(x)$ are fixed by a large deviation principle when $N\to \infty$.
The large deviation function can be derived following \cite{franz2019impact}. 
The corresponding saddle point equations can be written as
\beq
\tilde \qq_d(\omega_n)^{-1}= \frac 1\beta \frac{\omega^2_n}{\hbar^2} + \beta \mu + \Sigma(\omega_n)\ \ \ \ n=0,1,\ldots, \infty
\eeq
and the Lagrange multiplier $\mu$ is meant to be different from zero only for the spherical model where it is fixed self-consistently by the condition $\qq_d(0)=1$.
In the following we will analyze the RS phase and its instability towards a glassy, RSB, phase.
Since the model has a ${\mathbb Z}_2$ symmetry, the RS phase is characterized by $\QQ(x)=0$. Therefore we need to consider only $\qq_d(\hat t)$.
The self energy $\Sigma(\omega_n)$ admits an exact expression
\beq
\begin{split}
\Sigma(\omega_n)&=\frac{\alpha}{l_n}\hat \Sigma(\omega_n)\\
\hat \Sigma(\omega_n)&= -p_0^2\left(\frac{r}{1+r\tilde D(\omega_0)}\right)^2\frac{\delta\tilde D(\omega_0)}{\delta \tilde \qq(\omega_n)}\\
&+\sum_{m=0}^\infty l_m\frac{r}{1+r\tilde D(\omega_m)}\frac{\delta \tilde D(\omega_m)}{\delta \tilde \qq(\omega_n)}
\end{split}
\eeq
where
\beq
\begin{split}
\tilde D(\omega_p) &= \int_{0}^1 \de t D(t) e^{-i\omega_p t} \ \ \ D(\hat t) = \frac{\qq(\hat t)^2}{2}\:.
\end{split}
\eeq
The constant $r=\beta/\epsilon$ tunes the energy scale at a given temperature. Finally, $l_0=1$ and $l_{n>0}=2$ are multiplicity factors needed to enforce that $\tilde \qq_d(\omega_n)=\tilde \qq_d(-\omega_n)$.
The stability of this RS ansatz against RSB can be checked by looking at the condition
\beq
\l_R=1-\alpha \left(\frac{\tilde \qq(\omega_0) p_0 r}{1+r\tilde D(\omega_0)}\right)^2\geq 0\:.
\label{replicon_evalue}
\eeq
When the condition in Eq.~\eqref{replicon_evalue} is violated, one can easily show that the model undergoes to a continuous full replica symmetry breaking transition (as opposed to continuous transition towards 1RSB phases or RFOT type transitions \cite{SimpleGlasses2020}, which is the case of the mean field theory that would describe the models investigated in \cite{markland2011quantum, thomson2020quantum}). This can be checked by computing the Landau theory around the critical point and looking at the behavior close to the instability coming from the unstable phase. This is also what is found in the classical limit where $\hbar\to 0$, see \cite{urbani2022continuous} where the properties of the RSB phase have been investigated and solved.
Here we will not look at the details of the RSB phase. Rather we will be interested in exploring the phase diagram to see where one finds the emergence of glassy phases when quantum fluctuations are switched on.
It is natural to expect that strong quantum fluctuations tend to thermalize the system more than only classical thermal fluctuations.
However we are interested in investigating whether this phenomenology is monotonous in the strength of $\hbar$.

\begin{figure}
\includegraphics[width=\columnwidth]{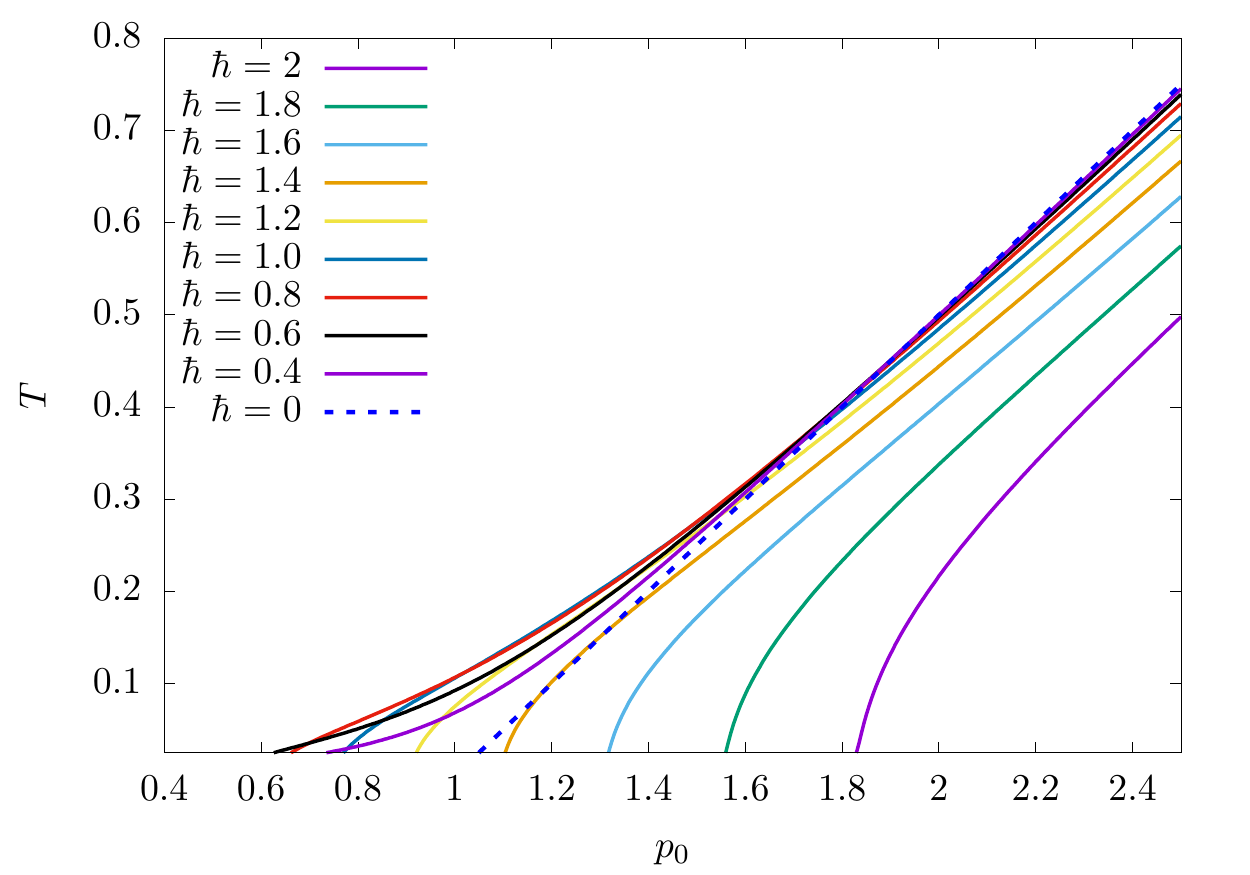}
\caption{The phase diagram for the spherical model. The lines denote the location of the RSB transition. At high temperature the model is RS, below the lines the model is glassy. The zero temperature landscape has a satisfiability transition at $p_0\simeq 1.8$ and a zero temperature RSB transition at $p_0=1$. Around $p_0\simeq 1.87$, small but finite $\hbar$ has deleterious effects for thermalization. The plot has been obtained by discretizing the thermal path integral with $2^{12}$ Trotter slices.}
\label{spherical_plot}
\end{figure}

\paragraph*{Results on the quantum models -- }
The RS saddle point equations can be solved numerically by truncating properly the Fourier series and using Fast Fourier Transform.
In the case of the spherical model, we can fix $\alpha=1/4$ and look for the phase diagram as a function of temperature, $\hbar$ and $p_0$. The free case where we do not impose the spherical constraint, has no fixed lenghtscale and therefore we can fix $p_0=1$ and change $\alpha$, $T$ and $\hbar$.
We start by presenting the results on the spherical model. 
In Fig.~\ref{spherical_plot} we plot the RSB transition lines in the $(p_0,T)$ plane for different values of $\hbar$.

The analysis of the classical model shows that there is a satisfiability transition around $p_0=p_0^*\simeq 1.87$, \cite{urbani2022continuous}. This means that at this point the canyon landscape disappears in the sense that for $p_0>p_0^*$ there are no configurations in phase space that satisfy all constraints $h_\mu=p_0$.
We clearly see that for $\hbar$ sufficiently large, quantum fluctuations manage to equilibrate the landscape in regions where thermal fluctuations would not be able to. This is essentially due to tunneling of energy barriers. However when we decrease $\hbar$, we observe that there is a crossing between the classical RSB line ($\hbar=0$) and the quantum RSB lines. This happens around the zero temperature satisfiability point for the classical potential energy landscape.
An extrapolation of the phase transition lines suggests that while the classical model is perfectly ergodic at $T=0$, the corresponding quantum model is not, if $\hbar$ is sufficiently small.
Therefore this suggests also that the $T=0$ RSB transition arrives before the classical one if $\hbar$ is moderately small. 
This somehow resembles what is found in another context, purely classical, where one tries to optimize a simple cost function over a disordered non-convex landscape \cite{sclocchi2022high}. 
Indeed one can think about the Suzuki-Trotter version of the path integral describing the quantum partition function and consider
first the limit $\epsilon\to 0$ and then $\beta\to \infty$.
In this case one is trying to accomodate an elastic closed chain of oscillators (a polymer) in the canyon landscape. This is nothing but the kinetic term of the quantum path integral. 
Therefore, one may expect that the landscape of the kinetic term of the quantum Hamiltonian,
given the canyon landscape, can become glassy before the classical RSB transition takes place.

\begin{figure}
\includegraphics[width=\columnwidth]{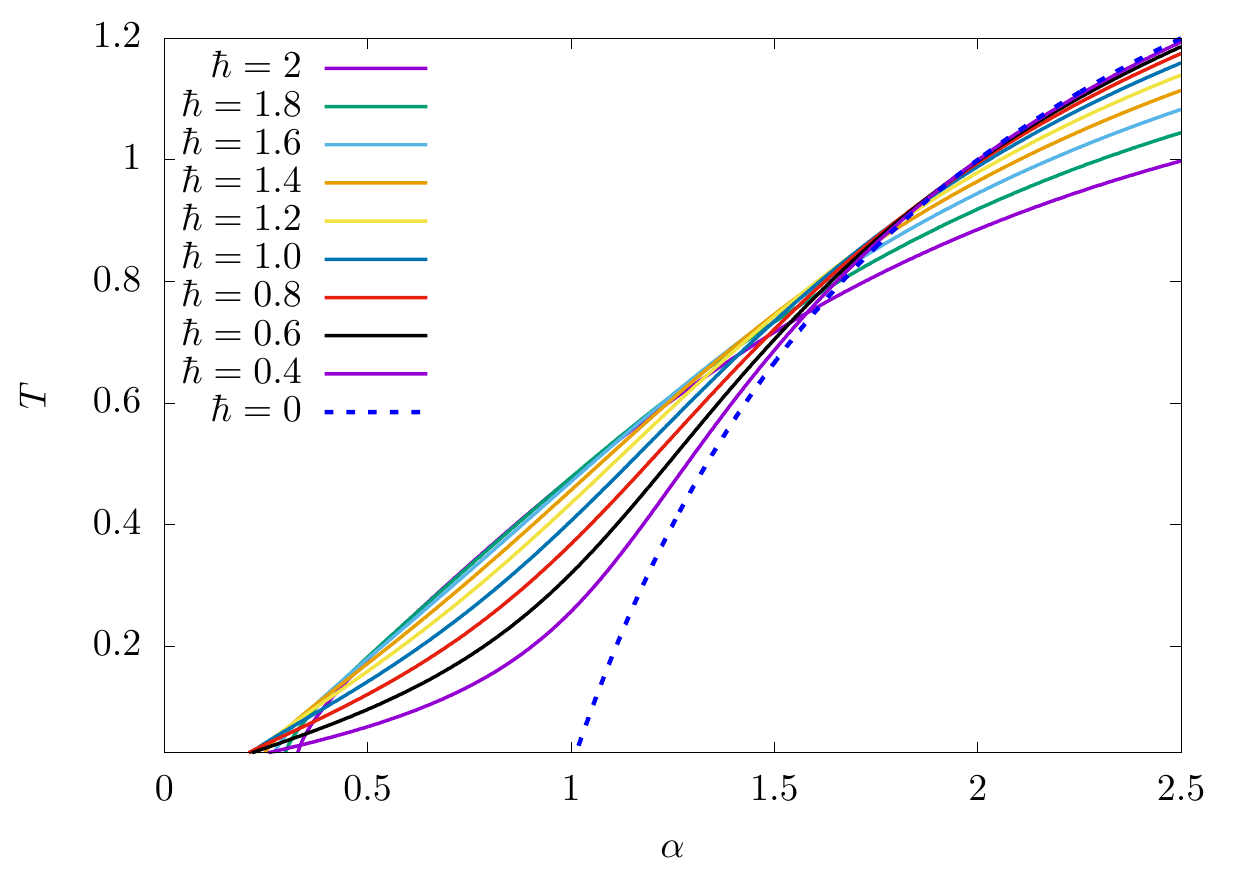}
\caption{We plot the RSB transition lines for the unconstrained model. The classical phase space is not confining when $\alpha<1/2$ and the classical RSB transition
happens at $\alpha=1$. The plot has been obtained by discretizing the thermal path integral with $2^{12}$ Trotter slices.}
\label{unconstrained_plot}
\end{figure}

We now turn to the study of the unconstrained model. 
We plot in Fig.\ref{unconstrained_plot} the corresponding phase diagram in the $(\alpha,T)$ plane for different values of $\hbar$.
Again we see that around the zero temperature glass transition of the classical model, a sufficiently small $\hbar$ has a deleterious effect to thermalization.
However here the most interesting effect can be seen at small $\alpha$. Indeed, the unconstrained model is not confining for $\alpha<1/2$ in the sense that the classical value of $\qq_d$ is divergent for $\alpha\to 1/2^+$. However, the numerical integration of the dynamical mean field theory equations shows that $q_d(0)<\infty$ which suggests that when quantum fluctuations are switched on, the quantum Gibbs measure gets picked around a hypersphere with a finite radius.

\paragraph{Conclusions and perspectives --} 
We have shown that in correspondence of satisfiability transition point where high-dimensional canyons disappear, moderate quantum fluctuations can have a deleterious effects to thermalization.

We expect that the quantum RSB phase for $T\to 0$, both above and below the classical satisfiability transition point, displays power law decaying correlations functions. In this phase one has a non-trivial $\QQ(x)$ and, in particular $\QQ_M>0$. Therefore we expect that if we look at $C(t)=q_d(t)-\QQ_M$, for $t\to \infty$ we should observe $t^{-\nu}$ with $\nu$ a critical exponent. A careful analysis of the RSB equations is mandatory to establish this phenomenology and compute the exponent $\nu$.
Furthermore, it would be very interesting to understand the effective field theory for the critical fluctuations on the same lines as in \cite{maldacena2016remarks} and to investigate correlation functions that are higher order than the bilinear investigated in this work \cite{chowdhury2022sachdev}. Both power-law decaying correlations and large fluctuations of response functions \cite{biroli2016breakdown} characterize the RSB phase which is marginally stable. It is very tempting to conjecture that the classical marginally stable potential energy landscape, once dressed with quantum fluctuations, provides a generic mechanism for the saturation of bounds to chaotic quantum dynamics. 
While, these aspects can be investigated in depth in the model defined by Eq.~\eqref{quantum_tis_mean}, it would be also very interesting to see how the mean field picture is changed in the finite dimensional version of the model described by Eq.~\eqref{Quantum_tissue}.

Finally we would like to point out that the analysis presented here will be very useful to study optimal control problems in high dimension. In \cite{urbani2021disordered} a prototype of these problems has been investigated mapping it to a quantum system. However the corresponding quantum impurity problem does not admit an exact solution and one needs to resort to quantum Monte Carlo. Therefore a possible perspective is to change the cost function in \cite{urbani2021disordered} in favor of the classical one analyzed in this work.
The classical optimal control problem would then be mapped to a quantum problem of the same kind of the one analyzed here, with two main differences: (i) in the optimal control case the trajectories are not periodic in (real) time and (ii) the off-diagonal part of the matrix $Q_{ab}(t,t')$ is time dependent. 
This would be useful to understand glassy phases of optimal control in high dimension and the competition between the entropy of paths joining points far in the landscape and the underlining structure of saddles at finite energy density.
It is clear that the mapping between the quantum model and the optimal control one provides an interesting route to address high-dimensional landscape properties. Quantum fluctuations, even at zero temperature, probe the non-local structure of the landscape and therefore may be used as lens to map it in detail.

%\bibliography{refs.bib}

\end{document}